# 八论以用户为中心的设计：一个智能社会技术系统新框架及人因工程研究展望

许 为 *

（浙江大学 心理科学研究中心，杭州 310058）

**研究要点：**
1. 分析传统社会技术系统（STS）理论在智能时代应用的局限性以及智能时代的 STS 新特征
2. 提出一个智能社会技术系统（iSTS）新框架，概括 iSTS 对智能系统研究和开发的意义
3. 针对今后 iSTS 的研究和应用，提出人因工程方法论和研究思路两方面的建议

**摘要** 传统的社会技术系统（STS）理论已经被应用在许多领域，但是智能时代的 STS 环境呈现出许多新特征，传统 STS 理论在智能时代的应用中表现出它的局限性。基于"以用户为中心设计"的理念，本文提出一个智能时代的智能社会技术系统（iSTS）新框架，并且概括了 iSTS 的新特征以及 iSTS 对在社会、组织环境中有效地开发和应用智能系统的意义。今后 iSTS 研究、设计、开发和应用的工作需要包括人因工程在内的跨学科协同合作，本文从人因工程方法论和研究思路两方面提出建议。

**关键词** 以用户为中心的设计，智能技术，社会技术系统，智能社会技术系统，人因工程
**中图分类号**：B849  **文献标识码**：A  **文章编号**：

## User-Centered Design（VIII）： A New Framework of Intelligent Sociotechnical Systems and Prospects for Future Human Factors Research

**XU Wei**

(Center for Psychological Sciences, Zhejiang University, Hangzhou 310058, China)

**Abstract:** Traditional sociotechnical systems (STS) theory has been widely used, but there are many new characteristics in the STS environment as we enter the intelligence era, resulting in the limitations of traditional STS. Based on the "user-centered design" philosophy and the perspective of human factors engineering, this paper proposes a new framework of intelligent sociotechnical systems (iSTS) and outlines the new characteristics of iSTS as well as its implications to the development of intelligent systems. Future research of iSTS requires interdisciplinary collaboration, including human factors engineering, this paper finally proposes suggestions from two aspects of human factors engineering methodology and approaches.

---

*作者：许为，博士，研究员；e-mail: xuweill@zju.edu.cn.

## 1. 引言

人因工程（human factors engineering）主要研究人机交互、人机界面设计以及它们对系统绩效的影响，不注重对宏观社会、组织环境等影响因素的考虑。进入智能时代，基于人工智能（AI）技术的智能系统在给人类带来益处的同时，也可能对人类的工作和生活产生负面影响。目前，一些研究者开始将智能系统研究从单纯的技术性研究转向包括社会、组织等因素的综合研究（Ågerfalk, 2020；Asatiani et al., 2021）。

社会技术系统（sociotechnical systems，简称 STS）理论由 Trist 及其英国塔维斯托克研究所的同事在 1950 年代提出。基于引入动力机器对工作、组织、管理、劳资关系、员工生活、员工家庭、煤矿工人协会等方面的影响，Trist 等人发现组织既是一个社会系统，同时又是一个技术系统，而技术系统对社会系统有很大的影响；个人态度和群体行为等方面都受到技术系统的重大影响 (Hoffman & Militello, 2008)。STS 理论提倡从社会和技术两方面来考虑技术的影响，通过实现社会、技术、组织等子系统之间的协调，实现社会和技术两方面的协同优化，从而达到最佳的整体系统绩效（Eason, 2011）。

任何一个新技术的推广使用都是在一定的 STS 环境中展开。进入智能时代，智能系统以及相应的自主化新特征使得 STS 变得更加复杂，采用 STS 理论的系统化理念有助于智能技术研发和推广使用（许为，2020；许为，葛列众，2018）。一方面，越来越多的智能系统对社会产生影响，例如，伦理道德，治理，社会文化等(Stahl, 2021)；另一方面，目前从 STS 角度来开展这方面的研究还远远不够（Borenstein et al., 2019；Behymer & Flach, 2016；Jones et al., 2013）。人因工程研究和设计通常只是针对 STS 中的一个部分：人机系统（Hodgson, et al., 2013），因此人因工程需要拓宽工作思路和范围，在宏观的 STS 视野中开展人因工程研究和应用（Dainoff et al., 2020），从而在社会与技术的宏观环境中优化人机系统的设计，最大限度地发挥新技术在社会和组织中的作用。

"以用户为中心的设计"理念强调系统开发要遵循"以人为中心"的流程和方法，从人类用户的需求出发，通过有效的"以人为中心"的人因工程等方法来优化人-机-环境系统的设计(许为，2003，2017)。STS 理论强调人机系统的设计既要考虑技术，而且要考虑处于社会、组织子系统环境中的人，优化人与技术的关系，强调"机器适应人"而不是"人适应机器"的解决方案，因此 STS 理论在本质上与人因工程学科分享"以用户为中心设计"的理念。

本文基于"以用户为中心的设计"理念，分析传统社会技术系统（STS)理论在智能时代应用的局限性以及智能时代所呈现的 STS 新特征，提出一个针对智能时代的智能社会技术系统（intelligent sociotechnical systems, 简称 iSTS）概念框架，目的是充实 STS 理论，助推智能系统的研究和应用。针对今后 iSTS 研究和应用，本文提出人因工程方法论和研究思路两方面的建议，进一步发挥人因工程在"以人为中心 AI"系统研发中的作用（许为，2019；Xu, Dainoff et al., 2022）。

## 2. 社会技术系统（STS）理论
### 2.1 基本概念

STS 包括相互独立但又相互依存的社会和技术两个子系统，其中，社会子系统一般包括人员(如知识、技能、态度、价值观和需求等)、角色、工作、文化和目标，涉及组织（如变革、结构、决策、奖励）等(Oosthuizen et al., 2019)；技术子系统一般包括物理基础设施、工具、技术和流程等 (Bostrom & Heinen, 1977)。STS 强调各个部分之间的协同，系统绩效依赖于技术子系统和社会子系统之间的协同优化和互补（Davis et al., 2014；Meijer & Bolívar, 2016；Pasmore et al., 2019），专注于 STS 中一个子系统而排除另一个子系统会导致系统整体绩效的下降 (Flach et al., 2012)。

STS 的目的是达到"双赢"的效果 (Pasmore et al., 2019)。引进新技术旨在促进社会、组织环境内新技术知识共享、学习和创新、优化设计等，实现人机协作和灵活性以获得竞争优势，更容易适应环境的变化，提升组织效益。但是，新技术可能会影响人们的工作和行为方式，可能导致他们的价值观、认知

结构、工作和生活方式等方面发生变化，而优化和可用的新技术有助于提升组织绩效和员工的工作满意度，最终提升人们的工作生活质量（Oosthuizen et al., 2019; Walker et al., 2008）。

然而，研究表明，技术和社会维度之间复杂的相互依存关系经常被忽视（Martin et al., 2018）。尽管 STS 理论的研究和应用已经开展了几十年，但是在一个大型复杂的社会环境中，STS 理论和方法对于许多行业（比如设计和系统工程界等）来说仍然相对较新（Norman & Stappers, 2015; Topi & Tucker, 2014）。近些年针对STS理论的研究兴趣日益浓厚，但并未得到广泛的重视 (Baxter & Sommerville, 2011)。

### 2.2 STS 理论对人因工程的影响

STS 理论扩宽了传统人因工程研究和应用的范围。传统人因工程通常强调在特定的环境中研究人机系统，以达到最佳的人-机-环境匹配。人因工程研究所考虑的环境通常是物理环境(照明、噪声、振动、温度、微重力等)（葛列众，许为，宋晓蕾，2022）。在过去的20年中，STS 已经开始影响了人因工程和工效学一些领域工作的发展，特别是对宏观工效学（Macroergonomics）领域发展的影响 (Waterson et al., 2015)。宏观工效学提倡在人机系统设计中需要在 STS 环境中设计整个工作系统，要考虑如何优化组织和管理等因素与技术之间的交互 (Hendrick & Kleiner, 2002)。Hendrick 和 Kleiner(2002) 定义宏观工效学为"一种自上而下的 STS 方法，用于分析和设计工作系统以及人-工作、人-机和人-软件界面的整体工作系统设计的应用"。宏观工效学还强调与 STS 理论分享系统开发的目标，追求 STS 中对社会系统（包括人、组织等）和技术子系统的协同优化，同时强调要优化各类"以人为中心"的界面，其中包括人机界面（硬件、设备等）、人-环境交互界面（环境照明、声音、空气质量等）、人-软件界面（软件人机界面、软件认知组件等）、人-组织环境界面（工作设计等）(Hendrik & Kleiner, 2000)。

STS 理论的系统化方法论影响了一些人因工程新型领域的发展。例如，Hollnagel & Woods（2005）提出协同认知系统理论，强调工作重点应该放在人机协同的合作认知系统的基础上，在系统设计中人因工程需要注重和优化人类操作员与技术系统之间的相互依赖关系。作为一门新兴领域，弹性工程（resilience engineering）基于 STS 理论从广泛的认知系统工程框架内关注组织中的安全和风险评估（Hollnagel, Woods, Leveson, 2006）。针对复杂 STS 系统，认知工作分析（CWA）从全工作领域的 STS 角度出发，为领域工作需求定义、分析提供了一组分析建模工具和一个概念框架，开始应用在各种工作领域中 (Vicente, 1999)。在应用上，人因工程研究开始将 STS 理论应用在健康医疗和计算系统安全等复杂领域的研究和系统开发中(Carayon, 2006; Stedmon et al., 2016)。STS 理论给人因工程的一个重要启示是，作为一门应用性学科，人因工程的研究不能局限于在实验室内闭门造车，需要在 STS 的大环境中充分考虑影响人因工程解决方案的各种因素。

STS理论也开始影响智能时代人因工程研究和应用的发展。进入智能时代，人因工程目前所对的研究对象是比以往任何时候更复杂的人-机-环境系统，比如智慧城市、智慧交通、智能制造、智能医疗等。智能时代的人机关系演化呈现出一系列人机交互新特征（比如人机协同合作关系），智能自主化技术也带来了一系列新特征和新问题（比如智能体的认知能力）（Xu，2020）。智能时代的这些新特征和新问题促使人因工程在更广阔的STS大环境中开展研究。 例如，智能系统设计需要充分考虑人类的各种新需求(如用户隐私、法律和伦理、决策权、技能成长等)，超越以往仅注重人机用户界面设计的点方案（point solution），追求端到端（end to end）的人因工程整体解决方案（许为，葛列众，2018，2020）。近几年，我们也试图从STS的角度来解决智能技术的一些人因问题。例如，许为（2019）提出的"以人为中心AI"的理念，以及许为，葛列众，高在峰. (2021) 所提出的人-AI交互（HAII）交叉学科新领域就是从宏观的STS环境中考虑；针对智能人机交互的研究，许为（2022a）采用协同认知系统理论分别分析了自动驾驶车人机共驾的协同认知生态系统；针对大型商用飞机驾驶舱单人飞行操作（SPO)的智能化系统设计，许为（2022b）提出了一个基于STS视野的针对SPO的协同认知生态系统框架。

由此可见，STS 理论拓宽了传统人因工程研究和应用的范围，影响了一些人因工程新型领域的发展，也开始影响智能时代人因工程研究和应用的发展。STS 理论的这些影响有助于人因工程学科超越传统研究和应用的思路和范围，从而能够从更广阔的社会、组织等角度来进一步优化人与新技术系统的交互和设计。然而，智能时代的 STS 环境呈现出许多新特征，传统 STS 理论在智能时代的应用中表现出它的局限性。为了更加有效地发挥 STS 理论对智能系统研究和应用的影响，STS 理论本身也需要进一步发展。

## 3. 智能社会技术系统（iSTS）
### 3.1 智能时代的 STS 新特征

进入智能时代，物联网智能网、智能社会、智能城市、智能交通等系统必将在各种因素相互依存的复杂社会技术生态系统中实现，这些智能系统本身就是基于智能技术的各种形态的 STS，它们的开发和使用也必将受到整个生态系统中各种因素的影响。

从 STS 技术子系统的角度看，如本系列文章（"五论以用户为中心的设计：从自动化到智能时代的自主化以及自动驾驶车"、"六论以用户为中心设计：智能人机交互的人因工程途径"）中所讨论的（许为，2020，2022a），智能时代的新型人机关系和智能人机交互新特征已经远远超出了传统人机交互研究和设计的范围。例如，智能技术以及所具有的自主化特征促进传统的计算技术从辅助人类工作的简单工具转变为人机合作伙伴式的复杂人机系统，由此形成了新型的人机关系 (Makarius et al., 2020；许为，2020)。这些新变化和新特征需要一种新思维和方法论来考虑如何更加有效地解决基于智能技术的人机交互等一系列新问题（许为，2022a）。从 STS 社会子系统的角度看，智能系统开发、引进和使用引发了一系列新的问题，比如 AI 伦理道德，用户需求（个人隐私，技能提升，决策权等），组织和员工对 AI 技术的可接受度，基于 AI 技术的组织决策等等。因此，智能时代的 STS 呈现出一系列新特征。表 1 概括地比较了智能与非智能时代 STS 之间的特征（许为，2020；许为，葛列众，2018，2020；许为，葛列众，高在峰，2021）。

表 1  智能与非智能时代社会技术系统的特征比较

| 特征项目 | 非智能时代的社会技术系统特征 | 智能时代的社会技术系统特征 |
| --- | --- | --- |
| 设计理念 | 以人为中心 | 以人为中心 |
| 系统组成 | 主要包括人类代理、机器、组织 | 主要包括人类代理、机器、组织以及具有认知特征的智能机器代理 |
| 认知代理 | 人类代理 | 人类代理,智能机器代理(单个或多重)(许为,2022a) |
| 机器角色 | 支持人类作业的辅助工具 | 支持人类作业的辅助工具,与人类合作的团队伙伴 |
| 核心技术 | 机械、计算机、数字自动化等 | AI 和机器学习、大数据、智能自主化技术等 |
| 人机关系 | 人机交互 | 人机交互 + 人机组队合作 (许为,葛列众,2020) |
| 用户需求 | 主要包括用户体验、心理、生理、安全等方面的需求 | 新增加的用户需求:个人隐私、伦理道德、情感、公平、决策权、技能成长等 |
| 系统输出可解释性 | 系统输出可解释性指系统输出信息为帮助用户理解而提供的解释程度,其主要取决于系统输出人机界面的可用性 | 可用性,AI"黑匣子"效应有可能导致系统输出难以解释,用户无法理解智能系统为什么、如何产生输出决定(许为,2019) |
| 系统决策和控制权 | 仅限于人类 | 人机决策分享,但必须保证人类拥有最终决控权;对智能系统决策透明度、不确定性、可解释性等方面的考虑(许为,2019) |
| 系统学习能力 | 单一的人类学习 | 人类学习,智能机器代理(机器学习等)的学习成长,人机共同学习和协同成长 |
| 系统设计范围 | 主要针对组织内部 | 整个系统内外生态系统 |
| 设计目标 | 基于一个相当静态固定的技术系统,强调社会子系统与技术子系统之间的优化;具有吸收有限的不确定性变化的能力 | 基于一个动态变化的技术系统,强调各部分(技术、组织、大数据信息、社会、外部生态系统等)之间的平衡优化;具有应对系统内外各种不确定性变化的能力 |
| 设计途径 | 通常由预算和项目时间计划、专家和管理层驱动,基于线性、自上而下的模式 | 迭代式原型化设计方法,数据和学习驱动式,基于经验和领域知识的混合式模式(包括自上而下、自下而上、内外部、自适应等) |
| 组织目标 | 高绩效,对变化的控制,相对稳定的系统 | 创新,应对动态变化的敏捷性和自适应能力 |
| 组织需求 | 工作系统设计(技术、流程、角色、人力资源、技能等),组织决策和计算辅助决策、组织设计和变革等 | 新增加的组织需求:新型人机关系驱动的人机合作和工作再设计,组织与智能系统之间的人机协同决策(包括决策权限设置,人类的决策权),针对智能自主化新特征的技术部署策略等 |
| 系统复杂性和开放性 | 相对独立和封闭的系统 | 更为复杂和开放的生态系统,各种形态的 STS 之间的交互和相互依存关系(智能物联网、智能社会、智能城市、智能交通等),智能系统的模糊边界和不可预测性,跨 STS 形态之间 AI 伦理道德的兼容性和交互影响等 |

如表 1 所列,相对于非智能时代的 STS,智能时代的 STS 呈现出一系列新特征和新问题。智能系统将促使人、机、社会、组织之间的交互和协同合作更紧密。面对复杂社会环境下更广阔范围内的问题与挑战,人机交互、人机智能协同等将以更加体系化的形式呈现(孙效华等,2020)。人因工程研究应该探讨如何有效地将智能系统嵌入到当前复杂的 STS 环境中,以渐进有效的方式促进人与复杂、智能化 STS 之间的融合。

## 3.2 iSTS 概念框架

综上所述，我们有必要进一步丰富传统的 STS 理论，这样才能有效地解决复杂智能系统的一系列新特征和新问题，从而发挥智能技术的最大优势和避免它对社会和人类带来潜在的负面影响。

已有研究者针对智能时代的 STS 理论和方法开展了一些初步研究。例如，Winby & Mohrman（2018）初步提出了一个针对数字化 STS 的设计方法，但是没有充分考虑智能技术的新特征。Steghofer 等（2017）认为下一代 STS 应该以智能技术为基础，STS 的社会子系统要涉及受 AI 影响的人类行为变化、AI 潜在的决策不确定性、复杂的社会组织因素（如组织、文化、道德、伦理等）等。也有人从工程实现和设计角度提出了针对智能时代 STS 的初步方案（Baxter & Sommerville, 2011; Jones et al., 2013）。在 Baxter & Sommerville（2011）的社会技术系统工程（STSE）框架中，他们将该框架建立在工作设计、信息系统、技术支持的人机合作和认知系统工程研究之上，试图弥合组织变革和系统开发之间的差距。针对智能时代 STS 的异构的、动态的、不可预测以及弱可控等特性，Dalpiaz 等人(2013) 提出了一种自适应 STS 的系统架构，试图通过执行监控-诊断-协调补偿循环的机制，使得 STS 具有自我重构的能力。

尽管已有一些初步研究，但是目前还没有完整的针对智能技术的 STS 框架，以往所提出的这些 STS 没有系统地考虑智能时代 STS 的新特征。针对智能技术和 STS 的新特征，本文初步提出一个智能社会技术系统（intelligent sociotechnical systems，简称 iSTS）的概念框架（见图 1）。

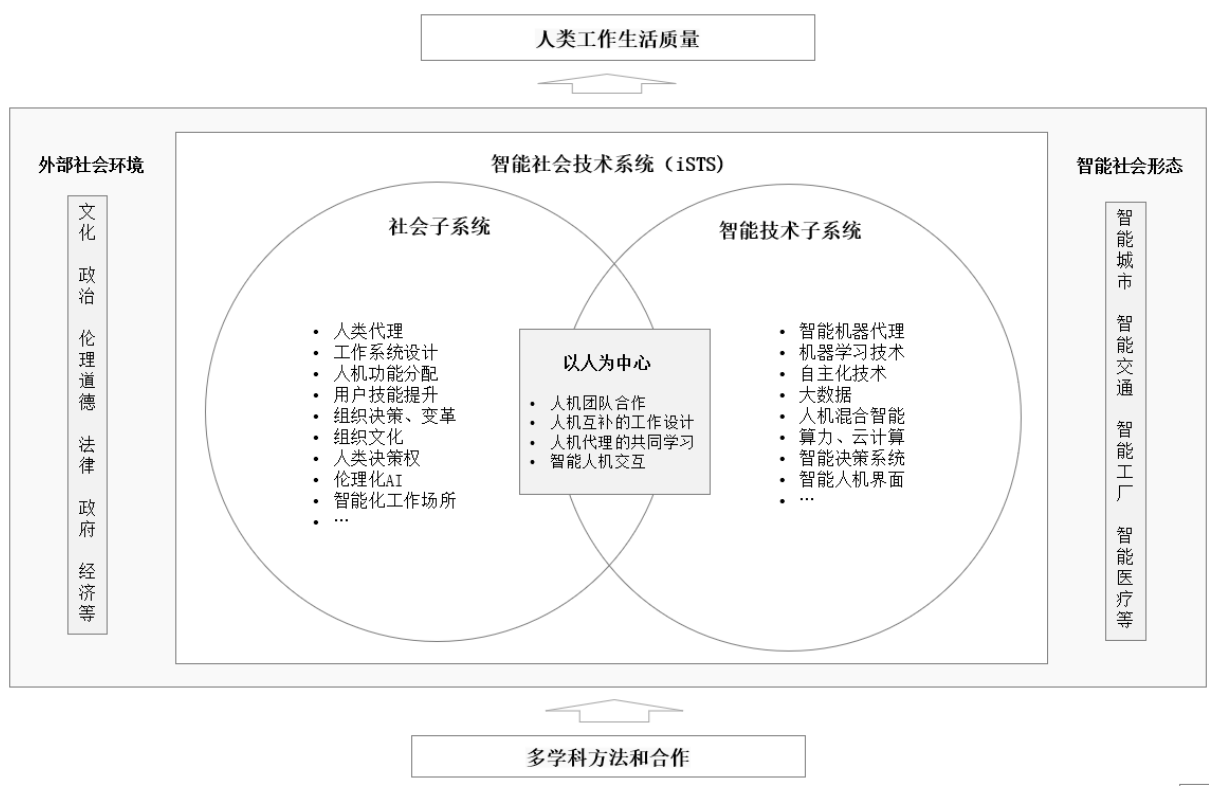

图 1 智能社会技术系统（iSTS）的概念框架

如图 1 所示，iSTS 继承了传统 STS 的一些基本特征。例如，iSTS 有一个内部环境（独立但相互依赖的技术和社会子系统），系统绩效依赖于技术和社会子系统的协同优化，专注于系统之一而排除另一个可能会导致系统绩效下降等（Badham et al., 2000）；iSTS 同样拥有一个宏观的外部环境以及各类智能社会形

态。但是，与传统 STS 相比，iSTS 有助于在社会、组织环境中有效地开发和应用智能系统，并且具有以下一些新特征。

首先，"以人为中心"的理念。一方面，基于"以人为中心"的理念，iSTS 将人以及相应的社会问题的解决置于优先地位，在实现智能技术与创新的同时，考虑教育、健康、交通等社会问题，通过社会技术生态的融合，解决 iSTS 中独特的人机合作、AI 伦理道德等新问题，实现智能系统开发和使用中人的参与、人类与智能系统的协作合作和有效交互，最终提高人类工作生活质量。另一方面，不同于传统 STS，iSTS 强调智能系统将改变传统的组织决策流程，AI、大数据等技术更加有效地辅助人类和组织的决策，也可能呈现人机决策分享等应用场景，但是系统设计需要定义决策权限等级，必须保证人类拥有最终决策操控权（Herrmann et al., 2018; Baxter & Sommerville, 2011）。

其次，人机团队协作。图 1 中社会子系统与智能技术子系统两个圆圈重叠的部分示意了两子系统之间的协同合作关系，这是 iSTS 区别于传统 STS 所特有的新型人机关系（许为，葛列众，2020）。在 iSTS 中，团队由人和智能系统机器组成，强调智能系统也是人机团队协同合作的成员，而不仅仅是传统 STS 中支持人类作业的简单工具，iSTS 中相互依赖的人机团队是分享共同目标的团体（Salas et al., 2008）。人类与智能系统（机器代理）之间的团队合作、互信、信息和决策分享等应该是 iSTS 成功的重要因素之一。

第三，人机优势互补的工作设计。实施 AI 新技术会改变既定的工作系统，可能导致用户陷入困境。iSTS 中的工作系统设计需要将人类和智能系统作为一个新型的工作系统重新设计，根据人与智能技术之间的优势互补，调整优化人机之间的功能和任务分配，其中包括人机角色分配、工作流程、作业环境等，从而有助于制定人机团队的工作和任务，保证人机功能的最佳分配，提升整体系统绩效。在引进智能技术的同时，要充分考虑员工的岗位重新分配、公平、技能成长等社会问题。iSTS 强调在智能系统设计中，需要人类的协同参与，将人类和技术无缝集成到优化的 iSTS 中（Behymer & Flach, 2016）。

第四，人机代理的共同学习。在复杂智能化社会环境中，iSTS 中社会和技术子系统之间存在复杂的相互作用和关系，它们跨越了传统的网络和物理、人和机器的界限，可以实现技术系统（硬件、软件和基于 AI 自主化技术的机器代理）和人类代理（在个人、社会和组织层面）之间的动态交互。智能代理是促进 iSTS 中社会和技术子系统之间动态交互的新资源，这种交互在短期内会调整智能代理自身的行为（基于机器学习算法等），并且会导致人类使用和期望模式的长期变化（社会学习）。同时，iSTS 的社会和技术子系统包含不同类型和层次的人类和智能代理自主权，从而使 iSTS 的学习成长、灵活性和自适应能力成为系统的基本特征。因此，iSTS 设计中必须考虑这种人机共同学习成长、不断共同发展的新特征，通过有效的设计和治理，这些代理的交互和合作可以为提高系统级别的灵活性和自适应能力创造新的机会（Heydari et al., 2019）。

第五，智能人机交互。基于 AI 技术的智能人机交互表现出一系列新特征，比如"人机双向式"交互，"人机协同合作式"交互，"情境化"交互，"多模态模糊推理式"交互等（许为，2022a）。一方面，基于智能人机交互、大数据等技术等的智能系统将有效地增强人机交互的自然性和有效性、系统的决策能力等；另一方面，相对于传统人机交互，智能系统会出现 AI "黑匣子"效应，导致系统输出难以解释和理解、决策透明度、不确定性、可解释性等方面的问题（许为，2019）。因此，iSTS 的设计开发需要考虑在充分发挥智能人机交互潜力的同时，也要避免智能技术给智能人机交互带来的潜在负面影响。

最后，开放式生态系统。在非智能时代，传统 STS 中的分析单元通常是一个组织或整个组织的有界部分，并且相对独立。在智能时代，物联网智能网、智能社会、智能城市、智能交通等各类 iSTS 的设计开发都存在于复杂的、相互依存的整个社会技术生态系统中，智能时代的 iSTS 与传统 STS 的区别之一在于智能机器代理的存在(Van de Poel, 2020)。在动态发展的社会环境中，智能代理的存在会增加，智能系统中的不确定性和不可预测性比传统 STS 系统更高，智能技术的认知学习、输出的不确定性等自主化特征带来了动态和模糊的 iSTS 边界。这些开放性特征给 iSTS 设计带来创新设计机遇的同时，也给系统设计、系统规则和价值观等方面带来了挑战（Hodgson et al., 2013）。因此， iSTS 的设计开发需要从一个开放式的人类、技术、社会、组织生态系统的角度考虑。

## 4. 人因工程对今后 iSTS 研发的学科支持

今后 iSTS 的研究、设计、开发和应用的工作需要包括人因工程在内的跨学科协同合作，作为交叉学科的人因工程可以从方法论和研究思路两方面提供学科支持。

### 4.1 人因工程的方法论支持

针对智能时代的 iSTS 研究和设计，研究者已经尝试一些方法。例如，Jones 等(2013) 强调针对社会子系统，建议采用抽象分析方法为计算框架的开发提供社会层面的信息，为 STS 的系统实施提供一个合适的平台。Kafali 等 (2019) 进一步提出了一个两层次的 STS 概念，由提供控制机制的技术层和表征利益相关者在规范方面需求的社会层所组成。不同于以往的方法，该研究采用智能代理作为计算实体，将社会维度纳入了实验验证过程。基于该模型的现场研究表明，该模型在一个智能代理环境中能够有效地模拟医院急诊场景的应用概念模式。针对复杂 STS 的研究，Tsvetovat & Carley, (2014) 认为需要一种综合的方法来解释这些系统内的心理学和社会学原理、交流模式和技术，并且采用多智能体系统的高保真模型方法，将分析模型与基于经验的模型相结合，从而为研究复杂的 STS 创建一个初步的模拟工具箱。

从 STS 开发流程和多学科合作的角度看，Fiadeiro (2008) 提出了"参与式多学科研究"的途径，Huang 等(2019) 进一步提出 STS 的设计流程应该是一个参与式的决策过程，需要用户、开发人员和其他利益相关者的积极参与，这样的流程能够保证新技术与社会和组织环境相匹配。本文提出的 iSTS 框架强调跨学科的团队合作，iSTS 的设计不是一个纯粹的技术项目，需要"以人为中心"的设计理念，通过人因工程迭代式的原型化和测试设计流程来逐步优化系统的设计和构建 (Norman & Stappers, 2015)。

近年来，人因工程中的许多领域采用了 STS 理论和方法(Waterson et al., 2015)，其中包括宏观工效学（Hendrick & Kleiner, 2002)、协同认知系统（Hollnagel & Woods, 2005)、弹性工程（Hollnagel, Woods, Leveson, 2006）等。这些理论和方法都试图在传统人因工程研究范围和方法的基础上考虑宏观的社会、生态、组织和技术等因素。国际工效学协会（IEA）定义了组织工效学，它涉及到与 STS 及其组织结构、政策和流程的优化，其中包含 STS 的许多方面，比如工作设计、团队设计、参与式设计、组织文化、虚拟组织和质量管理等 (Mumford, 2006; Baxter & Sommerville, 2011)。

针对今后 iSTS 的研发，表 2 概括了一些优先考虑的人因工程方法（Baxter & Sommerville, 2011; 许为，葛列众，2020)。表 2 同时还进一步列出了这些方法与 iSTS 设计开发流程阶段之间的关系，并且示意了这些方法在 iSTS 研发流程的各阶段中能够提供贡献的相对程度，其中"+ +"表示贡献程度高、"+"表示有贡献、"N/A"表示该方法一般不适合该阶段的工作。这些方法需要在今后 iSTS 的实践中进一步优化和充实。

### 表 2  人因工程类方法与 iSTS 开发阶段的关系

| iSTS 研究的主要人因工程方法 | 分析 | 设计 | 测评 |
| --- | --- | --- | --- |
| 以人为中心方法（ISO, 2010) | + + | + + | + + |
| 认知工作分析（Vicente, 1999) | + + | + | + |
| 协同认知系统（Hollnagel & Woods, 2005) | + + | + | + |
| 宏观工效学（Hendrick & Kleiner 2002) | + + | + | + |
| 人类学工作场所分析(Hughes et al, 1992) | + + | + | N/A |
| 情境设计（Beyer & Holtzblatt, 1999) | + + | + | N/A |
| 弹性工程（Hollnagel et al., 2006） | + | + | N/A |
| 纵向研究（Lieberman, 2009). | + + | N/A | + + |
| 生态研究方法（Brown et al., 2017) | + + | N/A | + + |

## 4.2 人因工程的研究思路

Dul 等人(2002) 在国际工效学联合会(IEA)的人因工程（工效学）发展战略建议中提出，人因工程的价值主张主要包含三个部分：(1)系统方法；(2)设计驱动；(3)绩效和福祉的设计目的。本文提出 iSTS 框架的目的是为了强调和优化人与新技术之间的相互关系，提升人类工作生活质量。针对今后 iSTS 的研究和应用，本文提出以下一些策略和建议。

首先，发挥学科优势。针对人因工程交叉学科的特点，美国人因和工效学会（HFES） 2016－2017 年度主席 Nancy Cooke（2017）总结了人因工程学科成功的三大要素，呼吁领域专业人员从三方面努力来发挥学科的作用：(1)从"以人为中心设计"理念出发，找准所要解决问题的切入点；(2) 从人-机-环境的角度出发，采用系统的思维和方法来全面考虑解决方案；(3) 充分考虑人因工程交叉学科的特点，加强与其他学科的协同合作。基于"以人为中心"理念、立足于多学科合作、站在系统层面上的的 iSTS 体现了这三大要素，因此开展 iSTS 研究和应用将有助于推动人因工程学科的进一步发展。

其次，分享"以人为中心"的理念。作为一门应用学科，人因工程必须能在应用中解决社会的实际问题。作为一门年青的学科，人因工程要进一步发挥学科影响力。美国人因和工效学会 (HFES) 2000－2001年度主席 William Howell（2001）提出了"分享理念"的模型，即人因工程需要与其他相近学科共同分享"以人为中心设计"的理念，而不是独自占有该理念的"封闭独享"模型。过去的 20 年，人因工程（包括工程心理学等）将该学科理念、方法和专业人才融入到人机交互、用户体验等领域的发展实践中，有力地促进了人机交互、用户体验等新领域的兴起和发展。这种实践充分体现了"分享理念" 模型的主张。在智能时代，"以人为中心的 AI"和"人-AI 交互"新领域的提出正是对 "分享模型"模型的进一步实践（Xu, 2019；许为，葛列众，高在峰，2021），而 Xu & Furie 等（2019）提出的交互、流程、集成和智能（interaction, process, integration, and intelligence, IPII）的人因工程新方法以及该方法在智能化供应链数字解决方案中的应用，以及本文提出的 iSTS 框架则是在更大的范围内分享和实践"以人为中心"的理念。因此，开展 iSTS 研究和应用将有助于进一步分享人因工程的"以人为中心"学科理念，解决更多的人类和社会的实际问题。

第三，坚持方法创新。人因工程研究和应用的方法要与时俱进（许为，2017，2018；许为，葛列众，2020）。iSTS 框架强调的是在智能时代开展人因工程研究和应用的一种设计新思维和新途径。针对 iSTS 框架的有效研究和应用实践需要创新的方法。近些年，人因工程涌现出一些新方法并且越来越多地被采用。例如，认知工作分析 (Vicente 1999)，弹性工程（Hollnagel et al., 2006），协同认知系统 (Hollnagel & Woods, 2005) 等方法从不同角度为分析和设计 STS 提供了有效的方法。作为实例，基于认知工作分析的认知工程方法对大型商用飞机自动化驾驶舱中人-自动化交互这一 STS 环境的分析获得了有效的研究结果（Xu, 2007）；针对自动驾驶车人机共驾和大型商用飞机驾驶舱单人飞行操作的协同认知生态系统框架为人因工程解决方案提出了一种新的工作思路（许为，2022a，2022b）。我们需要更多有效的能够系统化地分析和设计人、社会、组织以及技术因素的方法，帮助更好地了解人类、社会和组织因素如何影响开发和使用智能新技术，最终设计出优化的 iSTS。

第四，充实 iSTS 理论。本文所提出的仅仅是一个初步的 iSTS 概念框架，需要在今后的研究和应用中进一步充实。iSTS 框架本身还有许多问题需要解决。例如，在智能时代，在社会和技术子系统之间动态交互的基础上两者如何协同演化？人与智能系统的交互将如何影响社会和组织环境中的人类行为、组织变革、组织学习、组织认知等？如何有效地开展 iSTS 的集成设计和治理？另外，我们还可以借助其他的理论来丰富 iSTS 框架，比如采用基于协同认知系统理论的协同认知生态系统研究范式对自动驾驶人机共驾、大型商用飞机单人飞行操作的 STS 宏观环境的研究分析就是实例（许为，2022a；许为，2022b）。

最后，开展基于 iSTS 的人因工程研究和应用。人因工程研究和应用要超越传统的人机交互和人机界面设计的视野，在更广阔的 STS 视野中开展系统化的研究。我们面对的智能系统是复杂的，所处的 STS 环境也是复杂的。在人因工程研究和应用中如何运用 iSTS 还有许多问题需要解决。例如，如何将 iSTS 框架的理念有效地应用在智能系统的具体研发中？如何深入研究智能技术对人机交互、社会和组织的影响？尤其是智能系统对组织决策的影响（Pasmore et al., 2019；Keding, 2021）。iSTS 框架强调从宏观角度出发，

通过从文化、心理、组织、社会、技术等多学科角度来研究伦理化AI，人因工程如何为伦理化AI治理做出贡献（Chopra & Singh, 2018；Fiore, 2020）？未来智能社会形态正在形成，例如，"工业4.0"（Sony & Naik, 2020），"社会5.0"（周利敏，钟海欣，2019），网络物理系统（cyber-physical system, CPS）(Kant, 2016)，人因工程应该如何在智能时代发展传统的STS理论并且应用iSTS框架来作出学科贡献？

5. 总结

（1）人因工程通常强调人机交互和人机界面等方面的研究和设计，不注重宏观社会、组织环境等因素。进入智能时代，智能技术在给人类带来益处的同时，也可能对人类、社会、组织等方面带来影响。社会技术系统（STS）理论有助于从宏观的角度综合研究智能技术与各种因素之间的交互和协同优化。

（2）智能时代的STS呈现出一系列新特征和新问题。基于"以用户为中心"的理念，本文提出一个针对智能时代的智能社会技术系统（iSTS）概念框架。该iSTS框架突出了一些新特征，其中包括"以人为中心"的理念、人机团队协作、人机优势互补的工作设计、人机代理的共同学习、智能人机交互以及开放式生态系统。

（3）今后iSTS研究、设计、开发和应用的工作需要包括人因工程在内的跨学科协同合作。本文从人因工程方法论和研究思路两方面提出建议，希望通过充实智能时代的STS理论，开拓智能时代人因工程研究和应用的新思路。

13